\documentclass[letterpaper]{article}

\usepackage{cite, graphicx, amsmath, amsfonts, amssymb, array, latexsym, url, algorithm, algorithmic, multirow}

\newcommand{\rr}{{\bf r}}
\newcommand{\vv}{{\bf v}}
\newcommand{\uu}{{\bf u}}
\newcommand{\El}{\mathbb{L}_2}
\newcommand{\rrr}{\mathbb{R}^3}
\newcommand{\Es}{\mathbb{S}^2}

\begin{document}

\title{Spatially regularized compressed sensing of diffusion MRI data}

\author{Oleg~Michailovich, Yogesh~Rathi and Sudipto~Dolui
\thanks{O. Michailovich and S. Dolui are with the School of Electrical and Computer Engineering, University of Waterloo, Canada N2L 3G1 (e-mails: \{olegm,sdolui\}@uwaterloo.ca). Y. Rathi is with the Psychiatry Neuroimaging Laboratory (Department of Psychiatry, Brigham and Women's Hospital, Harvard Medical School), Boston, MA 02115 USA, (e-mail: yogesh@bwh.harvard.edu.)}}

\maketitle

\begin{abstract}
Despite the relative recency of its inception, the theory of compressive sampling (aka compressed sensing) (CS) has already revolutionized multiple areas of applied sciences, a particularly important instance of which is medical imaging. Specifically, the theory appears to provide an answer to the important problem of optimal sampling in MRI, with an ever-increasing body of works reporting stable and accurate reconstruction of MRI scans from the number of spectral measurements which would have been deemed unacceptably small as recently as five years ago. Reducing the number of MR measurements per scan comes to address one of the most critical impediments intrinsic in MRI, which is the relatively slow speed of image acquisition. Although very significant, such an improvement may still be insufficient in the cases when a repetitive acquisition of MRI scans pertaining to the same volume of interest is required. Acquisitions of this type are prevalent in diffusion MRI, in which an independent MRI scan is required to encode the strength of water diffusion along a predefined spatial direction. Thus, for example, an accurate delineation of multimodal diffusion profiles by means of high angular resolution diffusion imaging (HARDI) requires using  between 60 and 100 diffusion-encoding gradients. This, in turn, is translated into relatively long acquisition times, which adversely affects the applicability of HARDI for clinical diagnosing. To overcome this limitation, the present paper introduces a method for substantial reduction of the number of diffusion encoding gradients required for reliable reconstruction of HARDI signals. The method exploits the theory of CS, which establishes conditions on which a signal of interest can be recovered from its under-sampled measurements, provided that the signal admits a sparse representation in the domain of a linear transform. In the case at hand, the latter is defined to be spherical ridgelet transformation, which excels in sparsifying HARDI signals. What makes the resulting reconstruction procedure even more accurate is a combination of the sparsity constraints in the diffusion domain with additional constraints imposed on the estimated diffusion field in the spatial domain. Accordingly, the present paper describes a novel way to combine the diffusion- and spatial-domain constraints to achieve a maximal reduction in the number of diffusion measurements, while sacrificing little in terms of reconstruction accuracy. Finally, details are provided on a particularly efficient numerical scheme which can be used to solve the aforementioned reconstruction problem by means of standard and readily available estimation tools. The paper is concluded with experimental results which support the practical value of the proposed reconstruction methodology.  
\end{abstract}

%\begin{keywords}
%Diffusion MRI, HARDI, sparse representations, spherical ridgelets, total variation and compressed sensing.
%\end{keywords}

%------------------------------------------------ INTRODUCTION --------------------------------------------------
\section{Introduction}

The human brain consists of about $10^{11}$ nerve cells that can be subdivided into about 1000 different cell types, a complexity that far exceeds that of other organs of the body. A further complexity is evident in the way in which the component cells of the brain interconnect and function \cite{Jessell:1993zl}. In contrast to other types of the cells, each neuron communicates with many target cells by means of its distinctive protoplasmatic protrusion, called an axon. Axons with similar destinations, in turn, tend to form bundles - known as neural fibre tracts - which play a pivotal role in the determination of brain connectivity. Through reconstructing the pattern of connectivity of the neural tracts in both healthy and diseased subjects, it is therefore possible to obtain an abundance of valuable diagnostic information that could be used for early diagnostics of brain-related disorders, for assessing the damage caused to the brain by stroke, tumours or injuries, as well as for planning and monitoring of neurosurgeries \cite{Johansen-Berg:2009sh}.
 
Central to MRI is the notion of contrast, which is typically defined by the biochemical composition of interrogated tissue as well as by the morphology of its associated parenchyma. Prevalent in MRI practice are the contrasts determined by the T$_1$/T$_2$ relaxation times and proton density (PD). Despite their exceptional importance to clinical diagnosis, none of the above contrast mechanisms has demonstrated effectiveness in delineating the morphological structure of the white matter. It is only with the advent of diffusion MRI (dMRI) that scientists have been able to perform quantitative measurements of the diffusivity of white matter, based on which its structural delineation has become possible \cite{Bihan:1986lp,Basser:1994pd,Basser:1994ul,Pierpaoli:1996fu,Bihan:2001dp,Mori:2002ye,Johansen-Berg:2009sh}. Although over the last two decades dMRI has developed into an established technique with a great impact on health care and neurosciences, like any other MRI technique it remains subject to artifacts and pitfalls \cite{Bihan:1986lp}. While many of such artifacts can be overcome by using advanced hardware designs and/or more sophisticated imaging protocols \cite[Ch.2]{Johansen-Berg:2009sh}, \cite{Rohde:2004vn}, one particularly critical limitation of dMRI stems from the physics of the acquisition of diffusion MR images, and therefore is impossible to resolve by operational means. Specifically, since collecting the dMRI data requires a repetitive acquisition of MR responses from the same volume for a number of diffusion-encoding gradients, it is the relatively long acquisition times that greatly impair the practical value of this important imaging modality. Particularly, longer acquisition times entail a higher probability for the patient to exercise involuntary motion (typically caused by fatigue and/or stress related tremors, swallowing, uncontrollable sighing or coughing), which severely affects the quality of dMRI data. Moreover, since during the whole duration of the scan the patients are required to remain motionless, it is currently deemed ineffective to apply dMRI-based diagnosis in paediatrics as well as to patients with dementia or post-traumatic injuries, where non-compliance is typical. The problem of long acquisition times also hampers the application of dMRI for intra-operational imaging, where it could be an irreplaceable tool to use for neurosurgical planning and decision-making support \cite{Clark:2003ec,Akai:2005fk,Yu:2005fv}. Lastly, relatively long scanning times required by dMRI aggravate the problem of accessibility to MRI equipment. All the above arguments suggest that the practical value of dMRI could be improved by shortening the scanning times required for acquisition of dMRI data. A 
particular method to achieve such an improvement is detailed in this paper.

In this work, we adopt a general diffusion model in which each voxel within a region of interest (ROI) is allowed to support more than one fibre tract. In this case, under some general assumptions (see, e.g., \cite[Sect. 3.1]{Alexander:2005fk} for more details), the diffusion signal $s(\uu \, | \, \rr)$ originating from a voxel with spatial coordinate $\rr \in \rrr$ containing $M(\rr)$ fibres can be modelled as \cite{Tuch:2004cr,Alexander:2005fk}
\begin{equation}\label{multimodel}
s(\uu \,|\, \rr) =  s_0(\rr) \sum_{i=1}^{M(\rr)} \alpha_i(\rr) \exp \left\{ -b \, (\uu^T D_i(\rr) \, \uu) \right\},   
\end{equation}
where $\uu$ denotes the spherical coordinate, i.e.  
\begin{equation}\label{sphere}
\uu \in \Es : = \left\{ \vv \in \rrr \mid \|\vv\|_2 = 1 \right\},
\end{equation}
and $\alpha_i(\rr) > 0$ are positive weights obeying $\sum_{i=1}^{M(\rr)} \alpha_i(\rr) = 1$. In (\ref{multimodel}), $s_0$ denotes the diffusion signal obtained in the absence of diffusion encoding (i.e. the so-called ``b0 image"), $b$ is defined as a function of the shape and amplitude of diffusion-encoding gradients \cite[Eq. 3.18]{Mori:2007rw}, and $\{D_i(\rr)\}_{i=1}^{M(\rr)}$ are $3\times 3$ diffusion tensors associated with the $M(\rr)$ neural fibres passing through the coordinate $\rr$. In practical settings, the spherical coordinate $\uu$ is sampled at $K$ distinct points $\{\uu_k\}_{k=1}^K$ over the unit sphere. In this case, for each $\uu_k$, MR measurements are acquired in the form of a diffusion-encoded image $s_k(\rr) : =  s(\uu_k \,|\, \rr)$. As a result, a typical dMRI data set consists of a collection of such diffusion-encoded images $\{s_k(\rr)\}_{k=1}^K$, whose size $K$ determines the accuracy with which the directions of local diffusion flows can be estimated. 

Provided unlimited scanning time, one could measure the diffusion in thousands of orientations, making it possible to identify the directions of dominant diffusion with very high precision. For the reasons explained earlier, however, scanning times are always limited, which necessitates restriction of $K$ to a reasonably small value. This brings us to the central question addressed in the present paper: what is a sufficient number $K$ of diffusion-encoding directions to use? It turns out that, for some realizations of dMRI, the above question can be answered in a rigorous manner. In particular, in diffusion tensor imaging (DTI) \cite{Basser:1994pd,Basser:1994ul,Bihan:1986lp,Basser:2000lq,Bihan:2001dp,Mori:2002ye,Alexander:2007lh}, the reconstruction is carried out under the assumption that each voxel can support only one diffusion flow as most, which corresponds to setting $M(\rr) = 1$ for all $\rr$ in (\ref{multimodel}). Accordingly, a minimum of $K=7$ diffusion-encoded images are theoretically sufficient to measure $S_0(\rr)$ and recover the six non-repetitive components of the symmetric tensors $D(\rr)$ by means of least-square fitting. (In practice, however, a larger number of gradient directions is employed to reduce the estimation variance, with a typical $K$ being between 25 and 30 \cite{Jones:2004xd}.) Unfortunately, the  accuracy of DTI is known to deteriorate dramatically at the sites where the neural fibres (or bundles thereof) cross, touch upon each other, or diverge \cite{Wiegell:2000ij,Frank:2001mw,Alexander:2002uq,Frank:2002bd,Tuch:2002nx,Tuch:2003oq,Alexander:2005fk,Anderson:2005kx,Descoteaux:2006qf,Hess:2006lo}.

The fibre crossing problem in DTI has prompted efforts to develop dMRI methodologies which are capable of detecting multiple diffusion flows (or, equivalently, neural fibre tracts) within a given voxel. One of such techniques is High Angular Resolution Diffusion Imaging (HARDI)\cite{Frank:2001mw,Frank:2002bd,Tuch:2002nx,Tuch:2003oq,Tuch:2004cr,Hess:2006lo,Descoteaux:2006qf,Descoteaux:2007jw}, which is capable of capturing multi-modal diffusion patterns by sampling the spherical shell at a much greater number of points (usually between 60 and 100) as compared to the case of DTI. Increasing $K$ makes it possible to describe the diffusion measurements using much more accurate models. Among these are parametric models \cite{Kreher:2005mq,Banerjee:2005ys,McGraw:2006ov,Schultz:2008ys,Kaden:2008xz,Rathi:2009kx} which allow HARDI signals to be expressed in terms of a relatively small number of prototype signal forms. Unfortunately, fitting a parametric model often entails minimization of non-convex functionals, which is a noise-sensitive and computationally intensive task, prone to the problem of local minima. The need to predetermine the optimal number of fitting terms is known to be another disadvantage of using the models of the above type.

The problems associated with parametric modelling of HARDI signals can be overcome by using non-parametric models, in which case the signals are recovered by projecting the observed data onto properly defined functional subspaces. In particular, the applicability of spherical Fourier analysis \cite{Groemer:1996kx} to dMRI has been demonstrated in \cite{Anderson:2005kx,Descoteaux:2006qf,Hess:2006lo,Descoteaux:2007jw}, where HARDI signals are approximated by truncated series of spherical harmonics (SH). Despite its stability and computational efficiency, however, the SH-parameterization involves a relatively large number of SHs, which suggests that the SHs cannot be an adequate basis for sparse representation of HARDI signals. The main reason for this is rooted in the fact that the energies of elementary signals $d_i(\uu \,|\, \rr) := \exp \left\{ -b \, (\uu^T D_i(\rr) \uu) \right\}$ in (\ref{multimodel}) are concentrated alongside the great circles of $\Es$, whereas the energy of SHs is spread all over $\Es$, and, as a result, a relatively large number of SHs are needed to effectively ``encode" each of $d_i$. The inability of the basis of SHs to sparsely represent diffusion signals has led to the proposal of spherical ridgelets in \cite{Michailovich:2008gd,Michailovich:2010rp}, where it was shown that it only takes 6 to 8 spherical ridgelets on average to represent the HARDI signals with a precision exceeding the precision of their representation with 45 SHs.

The present work takes the ideas of \cite{Michailovich:2008gd,Michailovich:2010rp} one step further and shows that the availability of a sparsifying basis for HARDI signals can be used to reduce the number of diffusion gradients required for data acquisition. In particular, we suggest to use the theory of compressed sensing (CS) \cite{Candes:2006km,Candes:2006rw,Candes:2006dk,Donoho:2006mb,Donoho:2006cq,Haupt:2006hs,Baraniuk:2008zr} to recover the HARDI signals using the number of spherical samples $K$ in a range of values typical for DTI (i.e. $K \in [16, 24]$), thus allowing a multi-fibre analysis of dMRI data to be performed at the ``acquisition cost" of a standard DTI. 

It is worthwhile noting that the ideas of CS have already paved their way into the field of diffusion imaging \cite{Michailovich:2010xq,Landman:2010qq,Landman:2010pt,Menzel:2010hq,Lee:2010dw}. In this regard, conceptually the closest to the proposed approach is the method reported in \cite{Landman:2010qq}. In spite of this similarity, however, there are two principal distinctions which make the present method a more powerful alternative. In particular, the basis functions used in \cite{Landman:2010qq} are limited to represent an {\em average} diffusivity and anisotropy of the white matter, thereby neglecting both intra- and inter-voxel variability of tensors $D_i(\rr)$ in (\ref{multimodel}). The ridgelet representation, on the other hand, is a multiresolution technique, which possesses an intrinsic ability to deal with a continuum of different diffusion scales. Second, the approach in \cite{Landman:2010qq} is applied in a ``voxel-by-voxel" manner and it therefore does not take into consideration the spatial regularity of diffusion field. The present paper, on the other hand, proposes a novel formulation of the problem of CS-based reconstruction of diffusion signals, in which the sparsity constraints enforced in the diffusion domain are augmented by regularity constraints enforced in the spatial domain. The resulting reconstruction problem has the format of a convex minimization problem, which is solved using a specially adapted version of the split Bregman algorithm \cite{Yin:2008uf,Goldstein:2008yo,Esser:2008jo}. As will be shown below, the proposed algorithm results in a particularly advantageous computational structure which allows the solution to be computed via a sequence of simple and easily parallelizable steps.

The rest of the paper is organized as follows. Section II provides additional comments on the input-output structure of the proposed algorithm. The construction of spherical ridgelets is briefly outlined in Section III, whereas Section IV gives a formal description of the proposed reconstruction methodology. Some principal details on the numerical implementations of the proposed algorithm are summarized in Section V, with the results of our experimental studies reported in Section VI. Section VII finalizes the paper with a discussion and conclusions. 

%------------------------------------------------ DOS and DONTS --------------------------------------------------
\section{Problem Statement}\label{dosandonts}
In the centre of our considerations is the diffusion signal $s(\uu \,|\, \rr)$ which, when normalized by its related $b0$-image $s_0(\rr)$, quantifies the attenuation of MR readout caused by the diffusion of water molecules in the direction $\uu \in \Es$ through the spatial position $\rr \in \rrr$. In practical settings, both $\uu$ and $\rr$ are discretized. Specifically, restricting $\uu$ to a discrete set of orientations $\{\uu_k\}_{k=1}^K$ prescribes the acquisition of diffusion data in the form of $K$ diffusion-encoded images $\{s_k(\rr)\}_{k=1}^K$, with each $s_k(\rr)$ corresponding to a given $\uu_k$. In this case, for a fixed $\rr_0$, the vector $[s_1(\rr_0), s_2(\rr_0), \ldots, s_K(\rr_0) ]^T \in \mathbb{R}^K$ represents a discretization of $s(\uu \,|\, \rr_0)$. Note that such a discretization follows a linear measurement model, since each sample $s_k(\rr_0)$ can be represented as an inner product of $s(\uu \,|\, \rr_0)$ with a sampling function. In particular, let $\{\varphi_k(\uu)\}_{k=1}^K$ be a Dirac basis of sampling functions defined by
\begin{equation}\label{diracs}
\varphi_k(\uu) = \delta (1 - \uu \cdot \uu_k), \quad k=1,2,\ldots,K,
\end{equation}
where $\delta$ denotes the Dirac delta function and the dot stands for the Euclidean dot product. Then, formally,
\begin{equation}
s_k(\rr_0) = \langle s(\cdot \,|\, \rr_0), \varphi_k \rangle_{\El} := \int_{\Es} s(\uu \,|\, \rr_0) \, \varphi_k(\uu) \, d\eta(\uu), \,\, \mbox{ with } k = 1, 2, \ldots, K,
\end{equation}
with $d \eta$ being the standard rotation invariant measure on $\Es$.

Next, given a collection of $M$ spherical ridgelets $\{\psi_m(\uu)\}_{m=1}^M$ (defined below), the signal $s(\uu \,|\, \rr)$ is assumed to be expandable as 
\begin{equation}\label{represent}
s(\uu \,|\, \rr) = \sum_{m=1}^M c(\rr) \, \psi_m(\uu),
\end{equation}
with $c(\rr) \in \mathbb{R}^M$ being a vector of spherical ridgelet coefficients which depend on the spatial coordinate $\rr$. It is important to note that the set of spherical ridgelets is allowed to be overcomplete, implying $\dim \left[ {\rm Span}\{\psi_m(\uu)\}_{m=1}^M \right] < M$. A practical consequence of this fact is that the definition of coefficients $c(\rr)$ in (\ref{represent}) is, in general, not unique. This non-uniqueness is further aggravated by the fact that $c(\rr)$ will have to be recovered from an under-sampled set of diffusion measurements, in which case $K \ll M$.  Overcoming such a severe underdetermination in the problem of estimating the ridgelet coefficients $c(\rr)$ will be possible based on the fundamental premise of the theory of CS, which states that an accurate estimation of $c(\rr)$ is possible if the latter is sufficiently sparse {\em and} if the sampling and representation bases are sufficiently decorrelated. While the sparsity of $c(\rr)$  is rooted in the very design of spherical ridgelets \cite{Michailovich:2010rp}, the incoherency between the Dirac sampling functions (\ref{diracs}) and spherical ridgelets stems from the fact that  the former have an infinitely small support, whereas the latter are ``smeared" all over the unit sphere. The above properties of the basis of spherical ridgelets yield conditions for an effective application of CS, in which case one can obtain a faithful reconstruction of diffusion  signals using as few as $K=16$ diffusion-encoding gradients.     

The proposed algorithm produces an estimate of the ridgelet representation coefficients $c(\rr)$ in (\ref{represent}). Once available, the coefficients provide an access to the analytical definition of diffusion signals by virtue of (\ref{represent}). This can be used in a number of ways. One possibility could be to use the ridgelet coefficients to compute their associated orientation distribution functions \cite{Tuch:2004cr}, based on which a multi-fibre tractography analysis can be done \cite{Malcolm:2009pt,Malcolm:2009wa}. Alternatively, (\ref{represent}) can be used to evaluate the diffusion signals over an arbitrarily fine grid of orientations. Subsequently, such refined ``measurements" could be fitted using a different representation model, whose application to the original data would not have been possible without causing severe underestimation errors. Deconvolving the refined data to estimate the underlying fibre orientation functions \cite{Jansons:2003pr,Tournier:2004tw,Ramirez-Manzanares:2007fb,Sakaie:2007nx} would be another important option to follow. In this paper, we refrain from questioning which of the above possibilities is more advantageous over the others. Our sole objective here is to specify a signal processing algorithm which can be used to recover HARDI signals, while using the number of diffusion-encoded images typical for a standard DTI. 

Finally, it should be noted that the primarily purpose of the proposed methodology is to improve the value of HARDI in terms of its time efficiency. Since the improvement is achieved through merely decreasing the number of diffusion-encoding gradients, the proposed method by no means abrogates the use of fast imaging protocols \cite{Huang:2005fh,Jung:2007oa,Jung:2009le} to further accelerate the data acquisition. Furthermore, an additional speed-up can be achieved via applying CS to reconstruct the diffusion encoded images $s_k(\rr)$ from their subcritical samples in the spectral domain \cite{Lustig:2007bu,Kim:2009jb,Chartrand:2009xr,Trzasko:2009dq,Liang:2009ta}. Generally speaking, we believe this is a combination of such software and hardware technologies which will eventually lead to substantantial improvements in the practical value of HARDI-based diagnosing. In this paper, however, we confine our contribution to showing one particular way of attaining this important objective. 

%---------------------------------------------------- RIDGELETS ----------------------------------------------------
\section{Spherical ridgelets}
It is the property of spherical ridgelets to provide sparse representation of diffusion signals described by (\ref{multimodel}) which makes them an unparalleled tool for CS-reconstruction of HARDI data. To avoid repetitions, in what follows, we present only the most principal points of ridgelets design, while their detailed description can be found in \cite{Michailovich:2010rp}.

Spherical ridgelets are constructed using the fundamental principles of wavelet theory \cite{Daubechies:1995uq,Mallat:1999km}. Specifically, let $x \in \mathbb{R}_+$ and $\rho \in (0, 1)$ be a positive scaling parameter. Further, let $\kappa(x) = \exp \{ -\rho \, x \, (x + 1) \}$ be a Gaussian function, which we subject to a series of dyadic scalings to result in
\begin{equation}\label{kappaj}
\kappa_j(x) = \kappa(2^{-j} x) =  \exp \left\{ -\rho \, \frac{x}{2^j} \left( \frac{x}{2^j} + 1 \right) \right\},
\end{equation}
where $j \in \mathbb{N} : = \{0, 1, 2, \ldots \}$. Subsequently, the Gaussian-Weierstrass scaling function $\chi_{j,{\bf v}}: \Es \rightarrow \mathbb{R}$ at resolution $j \in \mathbb{N}$ and orientation ${\bf v} \in \Es$ can be defined as given by \cite{Freeden:1997kc,Freeden:1998ud}
\begin{equation}\label{WGF}
\chi_{j, \vv}(\uu) =  \sum_{n=0}^\infty \frac{2 n + 1}{4 \pi} \, \kappa_j(n) \, P_n( {\bf u} \cdot \vv), \quad \forall \uu \in \Es,
\end{equation}
where $P_n$ denotes the Legendre polynomial of order $n$. It is worth noting that the $\El$ energy of $\chi_{j, \vv}$ is concentrated around the spherical point $\vv$, with this concentration becoming more and more localized when $j$ approaches infinity. 

The spherical ridgelets are designed with the help of the Funk-Radon transform which, for an arbitrary continuous function $f: \Es \rightarrow \mathbb{R}$, is defined as
\begin{equation}\label{Funk}
\mathcal{R} \{f\} (\vv) = \int_{\uu \in \sigma(\vv)} f(\uu) \, \eta(\uu),  
\end{equation}
with $\sigma(\vv)$ denoting the great circle perpendicular to direction $\vv$, i.e. $\sigma(\vv) := \left\{ \uu \in \Es \mid \uu \cdot \vv = 0 \right\}$.  Subsequently, following \cite{Michailovich:2010rp}, the semi-discrete frame $\mathbb{U}$ of spherical ridgelets can be defined as
\begin{equation}\label{Uset}
\mathbb{U} : = \left\{ \psi_{j,{\bf v}} \mid {\bf v} \in \Es, j = -1, 0, 1, 2, \ldots  \right\},
\end{equation}
where the spherical ridgelet functions $\psi_{j, \vv}$ are obtained from $\chi_{j, \vv}$ according to  
\begin{equation}\label{above}
\psi_{j,{\bf v}} (\uu) = \frac{1}{2\pi} \mathcal{R}  \left\{ \chi_{j+1,\vv} - \chi_{j,\vv} \right\} (\uu),
\end{equation}
with $\chi_{-1,\vv}(\uu) \equiv 0$. Using (\ref{WGF}), the ridgelets (\ref{above}) can be redefined in a closed form as (see \cite{Groemer:1996kx} for details)
\begin{equation}\label{ridgelets}
\psi_{j, \vv}(\uu) = \frac{1}{2\pi} \sum_{n = 0}^\infty \frac{2 n + 1}{4\pi} \, \lambda_n \, \left(\kappa_{j+1}(n) - \kappa_j(n) \right) \, P_n({\bf u} \cdot {\bf v}),
\end{equation}
where $\kappa_{-1}(n) = 0, \forall n$ and 
\begin{equation}
\lambda_n =
\begin{cases}
2\pi {(-1)^{n/2}} \frac{1 \cdot 3 \cdots (n-1)}{2 \cdot 4 \cdots n}, \quad &{\rm if} \,\, n \,\, {\rm is \,\, even} \\
0, \quad &{\rm if} \,\, n \,\, {\rm is \,\, odd}.
\end{cases}
\end{equation}
 
The set $\mathbb{U}$ in (\ref{Uset}) is infinite-dimensional, and hence is not suitable for practical computations. To define a discrete counterpart of $\mathbb{U}$, one has first to restrict the values of the resolution index $j$ to a finite set $\left\{-1, 0, 1, \ldots, J \right\}$, where $J$ defines the highest level of ``detectable'' signal details. Additionally, the set of all possible {\em orientations} $\bf v \in \Es$ of spherical ridgelets needs to be discretized as well. To find a proper discretization scheme, we first note that the construction in (\ref{ridgelets}) suggests that the bandwidth of the  spherical ridgelets (and therefore the dimensionality of the functional space they belong to) increases proportionally to $2^j$. Since the space of spherical harmonics of degree $n$ has a dimension of $(n+1)^2$, it seems to be reasonable to define the number of ridgelet orientations at resolution $j$ to be equal to $M_j=  (2^{j+1} m_0 + 1)^2$, with $m_0$ being the smallest spherical order resulting in $\kappa_0(m_0) \le \epsilon$ for some predefined $0 < \epsilon \ll 1$ (e.g. $\epsilon = 10^{-6}$). Consequently, for each $j$, a total of $M_j$ orientations $\{\vv_j^i\}_{i=1}^{M_j}$ are chosen so that a discrete counterpart of $\mathbb{U}$ can now be defined as
\begin{equation}\label{ridges}
\mathbb{U}_d = \left\{ \psi_{j,\vv_j^i} \mid j=-1,0,1, \ldots J, \,\, i = 1,2,\ldots,M_j  \right\}.
\end{equation}
where the subscript $d$ stands for ``discrete". It should be noted that, although the set $\mathbb{U}_d$ is composed of continuously defined functions, its dimension is finite, since $\mathbb{U}_d$ consists of a total of $M = \sum_{j=-1}^J (2^{j+1} m_0 + 1)^2$ spherical ridgelets. To slightly simplify our notation, in what follows, the spherical ridgelets in $\mathbb{U}_d$ will be indexed as $\psi_m(\uu)$, with $m = 1, 2, \ldots, M$ being a combined index accounting for both different resolutions and orientations.

Given a sampling set of $K$ diffusion-encoding orientations $\{\uu_k\}_{k=1}^K$, one can use (\ref{ridgelets}) to compute the values of the spherical ridgelets in $\mathbb{U}_d$ over the sampling set\footnote{Since the definition in (\ref{ridgelets}) involves an infinite summation, the latter needs to be truncated to render the computations practical. In practice, we truncate the summation to index $n_{max}$ for which the magnitude of the summand drops below $10^{-9}$.}. The resulting values can be stored into a $K\times M$ matrix $A$ defined as
\begin{equation}\label{matrix}
A = 
\left[ \begin{array}{cccc}
\psi_1(\uu_1) & \psi_2(\uu_1) & \ldots & \psi_M(\uu_1) \\
\psi_1(\uu_2) & \psi_2(\uu_2) & \ldots & \psi_M(\uu_2)\\
\ldots & \ldots & \ldots & \ldots \\
\psi_1(\uu_K) & \psi_2(\uu_K) & \ldots & \psi_M(\uu_K)
\end{array} \right].
\end{equation}
Then, for a given vector $s(\rr) := \left[ s(\uu_1 \,|\, \rr), s(\uu_2 \,|\, \rr), \ldots, s(\uu_K \,|\, \rr)  \right]^T \in \mathbb{R}^K$ of the measured values of a diffusion signal $s(\uu \,|\, \rr)$ at the spatial location $\rr$, the model (\ref{represent}) asserts the existence of representation coefficients $c(\rr) \in \mathbb{R}^M$ such that
\begin{equation}\label{noisymodel}
s(\rr) = A\,c(\rr) + e(\rr), 
\end{equation}
where $e(\rr)$ accounts for both model and measurement noises. Clearly, the non-negligibility of $e(\rr)$ along with the fact that $K \ll M$ makes the problem of recovering the representation coefficients $c(\rr)$ from $s(\rr)$ a very challenging inverse problem, our solution to which is presented next.

%---------------------------------------------- RECONSTRUCTION ------------------------------------------------ 
\section{Proposed reconstruction framework}
Let $\Omega$ represent the volume within which diffusion measurements are acquired. Also known as a region of interest, $\Omega$ is assumed to be a bounded rectangular subdomain of $\rrr$, i.e. $\Omega : = [0, L_x] \times [0, L_y] \times [0, L_z] \subset \rrr$. Let $\Omega_d$ be a discrete subset of $\Omega$, which represents the spatial locations at which the diffusion signal is measured. Specifically, $\Omega_d$ is assumed to be a uniform lattice which can be formally defined as
\begin{equation}\label{set}
\Omega_d := \left\{ \rr = \{x_{i_1},y_{i_2},z_{i_3}\} \, \Big| \, x_{i_1} = \frac{i_1}{N_x} L_x, \, y_{i_2} = \frac{i_2}{N_y} L_y, \, z_{i_3} = \frac{i_3}{N_z} L_z \right\}, 
\end{equation}
where $0 \le i_1 < N_x$, $0 \le i_2 < N_y$, and $0 \le i_3 < N_z$ are sampling indices in the direction of $x$, $y$ and $z$ coordinates, respectively.

Let $K$ be the number of diffusion-encoding gradients used for HARDI data acquisition, and let the corresponding gradient orientations be denoted by $\{\uu_k\}_{k=1}^K$, where $\uu_k \in \Es$. For each of these values of $\uu_k$, MRI measurements result in its associated diffusion-encoded image $s_k$, which can be formally viewed as a mapping from $\Omega_d$ to $\mathbb{R}$. For the sake of convenience, each $s_k$ can be stored and manipulated as an $N_x\times N_y\times N_z$ array of real numbers, namely $s_k \in \mathbb{R}^{N_x\times N_y\times N_z}$. Alternatively, at a given coordinate $\rr \in \Omega_d$, one can combine the values $s_1(\rr), s_2(\rr), \ldots, s_K(\rr)$ into a column vector $s(\rr):=[s_1(\rr), s_2(\rr), \ldots, s_K(\rr)]^T \in \mathbb{R}^K$ (as it was already done in (\ref{noisymodel})). This vector can then be regarded as a vector of discrete measurements of an associated HARDI signal $s(\uu \,|\, \rr)$ measured at orientations $\{\uu_k\}_{k=1}^K$. It is worth noting that, according to the above notations, the value $s_k(\rr)$ admits a twofold interpretation, {\em viz.} either as the $k^{\rm th}$ coordinate of vector $s(\rr)$ or as the value of image $s_k$ at spatial position $\rr$.

When combined together, the continuum of vectors $s(\rr)$ can be regarded as a discrete vector field $s: \Omega \rightarrow \mathbb{R}^K$, in which case $s(\rr)$ has a natural interpretation of the value of $s$ corresponding to position $\rr$. The vector space $\mathfrak{V}$ of such vector fields can be endowed with the standard inner product
\begin{equation}\label{inner}
\langle s^1, s^2 \rangle_\mathfrak{V} = \sum_{\rr \in \Omega_d} s^1(\rr)^T s^2(\rr) = \sum_{\rr \in \Omega_d} \sum_{k=1}^K s_k^1(\rr) \, s_k^2(\rr) = \sum_{k=1}^K \langle s_k^1, s_k^2 \rangle,  
\end{equation}
with $\langle s_k^1, s_k^2 \rangle = \sum_{\rr \in \Omega_d} s_k^1(\rr) \, s_k^2(\rr)$ being the standard inner product between the scalar-valued images $s_k^1$ and $s_k^2$. Accordingly, congruent to the definition in (\ref{inner}), the $\ell_2$-norm of $s \in \mathfrak{V}$ is defined as  
\begin{equation}\label{ell2}
\| s \|_{\mathfrak{V}, 2} = \Big[ \sum_{\rr \in \Omega_d} \|s(\rr)\|_2^2 \Big]^{1/2}= \Big[ \sum_{k=1}^K \| s_k \|_F^2 \Big]^{1/2},
\end{equation}
where $\| \cdot \|_2$ and $\| \cdot \|_F$ denote the Euclidean vector and Frobenius matrix norms, respectively.

Another norm on $\mathfrak{V}$ that we shall make use of is the total variation (TV) semi-norm which is defined as follows. First, let us define the total variation of the $k^{\rm th}$ component $s_k$ of the field $s$ in a standard manner as
\begin{equation}
\|s_k\|_{\rm TV} = \sum_{\rr \in \Omega_d} \Big[ \sum_{{\bf p} \in \mathcal{C}(\rr)} \left| s_k(\rr) - s_k({\bf p}) \right|^2 \Big]^{1/2},
\end{equation}
where $\mathcal{C}\left(\rr = (x_{i_1}, y_{i_2}, z_{i_3})\right) = \left\{ (x_{i_1-1}, y_{i_2}, z_{i_3}), (x_{i_1}, y_{i_2-1}, z_{i_3}), (x_{i_1}, y_{i_2}, z_{i_3-1}) \right\}$ is a 3-neighbourhood (causal) clique of voxel $\rr$. Consequently, the TV norm of the discrete vector field $s$ can be defined in terms of the TV-norms of its $K$ components as
\begin{equation}\label{TV}
\| s \|_{\mathfrak{V},{\rm TV}} = \Big[ \sum_{k=1}^K \|s_k\|_{\rm TV}^\alpha \Big]^{1/\alpha}.
\end{equation}
Thus, for example, $\alpha = 2$ was used in the TV-denoising method reported in \cite{Blomberg:1998fk}. In this paper, we use $\alpha = 1$.

Now, let $\{ \psi_m \}_{m=1}^{M}$ be a set of spherical ridgelets defined by (\ref{ridges}), which is assumed to be rich enough so that each HARDI signal can be expressed according to (\ref{represent}). Analogously to the discrete measurements $s(\rr)$, the representation coefficients $c(\rr)$ corresponding to different voxels $\rr$ can be aggregated into a vector field $c \in \mathfrak{U}$, where $\mathfrak{U}: \Omega_d \rightarrow \mathbb{R}^M$ (with $c(\rr)$ being the value of $c$ observed at $\rr$). Although it is possible to endow the vector space $\mathfrak{U}$ with both the $\ell_2$- and TV-norms by analogy with (\ref{ell2}) and (\ref{TV}), it will be particularly useful to consider the $\ell_1$-norm of $c$ which can be defined as
\begin{equation}
\| c \|_{\mathfrak{U}, 1} = \sum_{\rr \in \Omega_d} \| c(\rr) \|_1 = \sum_{\rr \in \Omega_d} \sum_{k=1}^M | c_k(\rr) |.
\end{equation}

Using the definitions of the vector fields $\mathfrak{V}$ and $\mathfrak{U}$ as well as the definition of $A$ in (\ref{matrix}), a connection between $\mathfrak{V}$ and $\mathfrak{U}$ is established by means of a linear map $\mathcal{A}: \mathfrak{U} \rightarrow \mathfrak{V}$ that is defined as given by
\begin{equation}
\mathcal{A}: \mathfrak{U} \rightarrow \mathfrak{V}: c(\rr) \mapsto s(\rr) = A \, c(\rr), \, \forall \rr \in \Omega_d.
\end{equation}
Consequently, using $\mathcal{A}$, one can define the HARDI data formation model as
\begin{equation}\label{data}
s  = \mathcal{A}\{c\} + e, 
\end{equation}
where $e \in \mathfrak{V}$ is supposed to account for both measurement noise and modelling errors, and it is assumed to have a relatively small $\ell_2$-norm $\|e\|_{\mathfrak{V},2} \le \epsilon$.

The model (\ref{data}) suggests a reduction of the problem of estimation of HARDI signals to the problem of estimation of their corresponding representation coefficients $c$ from the discrete and noisy measurements $s$. Furthermore, as our main intension has been to recover the coefficients $c$ using as few diffusion-encoding gradients as possible (implying $K \ll M$), there is an infinite number of solutions which would fit the constraint $\| \mathcal{A}\{c\} - s \|_{\mathfrak{V},2} \le \epsilon$. However, if it is known {\em a priori} that, for each $\rr \in \Omega_d$, the vector of representation coefficients $c(\rr)$ is sparse, then a useful estimate of $c$ can be obtained as a solution to the following convex optimization problem \cite{Candes:2006km,Candes:2006rw,Candes:2006dk,Donoho:2006mb,Donoho:2006cq,Haupt:2006hs,Baraniuk:2008zr}
\begin{align}\label{L11}
&\min_c  \| c \|_{\mathfrak{V}, 1} \\
\mbox{s.t.  } \| \mathcal{A}&\{c\} - s \|_{\mathfrak{V},2} \le \epsilon. 
\end{align}
It should be noted that the optimization problem (\ref{L11}) is equivalent to solving
\begin{align}
\min_{c} &\sum_{\rr \in \Omega_d} \| c(\rr) \|_1 \\
\mbox{s.t.  } \sum_{\rr \in \Omega_d} \|A \, c&(\rr) - s(\rr) \|_2^2 \le \epsilon^2, 
\end{align}
and therefore, under the assumption of spatially homogeneous noise $e$, the problem (\ref{L11}) is separable in the spatial coordinate $\rr$. This means that an optimal field $c$ can be recovered by solving for its components
\begin{align}\label{L12}
\min_{c(\rr)}  &\| c(\rr) \|_1 \\
\mbox{s.t.  } \| \mathcal{A}\{c(\rr)\} - s(\rr) &\|_2 \le (N_x N_y N_z)^{-1/2} \epsilon,
\end{align}
{\em independently} at each $\rr \in \Omega_d$. 

While computationally attractive, the above solution is suboptimal, since it completely disregards the dependencies which are likely to exist between spatially adjacent HARDI signals. A possible way to take such dependencies into consideration is to require the noise-free version of the measured signal $s$ to possess a minimal TV norm among all possible candidate solutions \cite{Rudin:1992fh}. This requirement can be translated into the following minimization problem 
\begin{align}\label{cost}
\min_c  \Big\{ \| c \|_{\mathfrak{V}, 1} + &\gamma \, \| \mathcal{A}\{c\}\|_{\mathfrak{V},{\rm TV}} \Big\} \notag \\
\mbox{s.t.  } \| \mathcal{A}\{c\} &- s \|_{\mathfrak{V},2} \le \epsilon. 
\end{align}
where the role of $\gamma > 0$ is to balance the relative influence of the sparse and TV terms in the above cost function. The optimization problem (\ref{cost}) can be rewritten in its equivalent Lagrangian form
\begin{equation}\label{L-cost}
\min_c  \left\{ \frac{1}{2} \| \mathcal{A}\{c\} - s \|_{\mathfrak{V},2}^2 + \lambda \, \| c \|_{\mathfrak{U}, 1} + \mu \, \| \mathcal{A}\{c\}\|_{\mathfrak{V},{\rm TV}} \right\},
\end{equation}
for some optimal values of $\lambda > 0$ and $\mu > 0$ \cite{Boyd:2004fy}. 

Below, we are going to specify a particular, computationally efficient method for solving (\ref{L-cost}). In this connection, it is instructive to outline the following two instances of (\ref{L-cost}).

\subsection{Sparse-only reconstruction}\label{SP-only} When $\mu = 0$, the functional in (\ref{L-cost}) becomes {\em separable in the spatial variable} $\rr$ in the sense that, in such a case, an optimal $c$ can be recovered by solving
\begin{equation}\label{L1only}
\min_{c(\rr)} \left\{ \frac{1}{2} \| A \, c(\rr) - s(\rr) \|_2^2 + \lambda \, \| c(\rr) \|_1 \right\}
\end{equation}
for each $c(\rr)$ independently. Note that (\ref{L1only}) can be considered to be a Lagrangian form of the optimization problem (\ref{L12}). There exist a broad spectrum of methods which could be used for solution of (\ref{L1only}). Some particularly attractive algorithms seem to be those exploiting the principle of iterative shrinkage (aka iterated thresholding) \cite{Daubechies:2003bv,Figueiredo:2003qc,Elad:2007ss,Bioucas-Dias:2007bs}. While the non-differentiability of $\ell_1$-norm in (\ref{L1only}) rules out the applicability of gradient-based optimization tools, iterative shrinkage methods are capable of finding a solution of (\ref{L1only}) through iterative application of a first-order, fixed-point update rule. Specifically, many algorithms of this kind find an optimal solution as a stationary point of the sequence of estimates produced by
\begin{equation}\label{ista}
c(\rr)^{t+1} = \mathcal{S}_{\lambda/\nu} \left\{ c(\rr)^t + \nu^{-1} A^T \left( s(\rr) - A \, c(\rr)^t \right) \right\},
\end{equation}
where $ \mathcal{S}_\tau\{ t \} = {\rm sign}(t) (| t | - \tau)_+$ denotes the operator of soft thresholding and $\nu$ is chosen to obey $\nu > \| A A^T\|$. In the present paper, a modification of the iterative update in (\ref{ista}), known as the {\em fast iterative shrinkage thresholding algorithm} (FISTA) \cite{Beck:2009fx}, was employed due to its considerably faster convergence as compared with many alternative ``accelerated" methods.  

It should be emphasized that, while being suboptimal from the viewpoint of spatial-domain regularity, the solution of (\ref{L1only}) through iterative shrinkage is advantageous in two important practical ways. First, it suggests considerable storage reduction, since the thresholding operator in (\ref{ista}) sets to zero the representation coefficients with amplitudes less or equal to $\lambda / \nu$ in absolute value. It makes it possible to use sparse data formats to store and manipulate the representation coefficients. Second, the fact that the estimation of $c(\rr)$ is performed at each voxel independently suggests a natural way to speed up the overall estimation process though parallel computing on a multicore system. 

\subsection{TV-only reconstruction}\label{TV-only} When $\lambda = 0$, solving the optimization problem (\ref{L-cost}) is equivalent to simultaneously solving $K$ optimization problems of the form
\begin{equation}\label{TV-1}
\min_c \left\{ \frac{1}{2} \| [\mathcal{A}\{c\}]_k - s_k  \|_F^2 + \mu \, \| [\mathcal{A}\{c\}]_k \|_{\rm TV}  \right\},
\end{equation}
where $k=0,1,\ldots, K-1$ and $[\mathcal{A}\{c\}]_k$ denotes the $k$-th component of the vector field $\mathcal{A}\{c\} \in \mathfrak{V}$. Let $[\mathcal{A}\{c\}]_k$ be denoted by $u_k$, i.e. $u_k := [\mathcal{A}\{c\}]_k$. Then, reformulated with respect to $u_k$, the problem (\ref{TV-1}) can be rewritten as
\begin{equation}\label{TV-2}
\min_{u_k} \left\{ \frac{1}{2} \| u_k - s_k  \|_F^2 + \mu \, \| u_k \|_{\rm TV}  \right\},
\end{equation}
in which case it can be recognized as the problem of TV-denoising of the diffusion-encoded image $s_k$ \cite{Rudin:1992fh}. It is important to note that, in contrast to (\ref{TV-1}), the problem (\ref{TV-2}) can be solved for each $k$ independently, in which case we say that the estimation becomes {\em separable in the diffusion direction}.

The current arsenal of methods which can be used for solving (\ref{TV-2}) is broad. Originated from the work of Rudin {\em et al} \cite{Rudin:1992fh}, TV-denoising methods currently include gradient-based optimization methods \cite{Vogel:1996dz}, first-order methods \cite{Aujol:2009fu}, iterative shrinkage methods \cite{Wang:2010fk, Michailovich:2010uq}, and Bregman-type iterative algorithms \cite{Osher:2005ye,Yin:2008uf,Goldstein:2008yo}. The first-order methods appear to be a particularly attractive option when relatively large data arrays need to be processed, which is the case relevant to dMRI. Consequently, in the present paper, we employ the algorithm of \cite{Chambolle:1997qo} for the simplicity and elegancy of its implementation as well as for its outstanding convergence properties. 

%----------------------------------------------------- BREGMAN -----------------------------------------------------
\section{Solution using Split Bregman Algorithm}\label{BREG}
Directly solving the original problem (\ref{L-cost}) is difficult because of the compound nature of the regularization it involves. The split Bregman approach \cite{Yin:2008uf,Goldstein:2008yo} allows reducing (\ref{L-cost}) to a simpler form through introduction on an auxiliary variable $u \in \mathfrak{V}$, which can be viewed as a noise-free version of the data field $s$. Particularly, using $u$ (\ref{L-cost}) can be redefined as
\begin{align}\label{B-cost}
\min_{c, u}  \Big\{ \frac{1}{2} \| u - s \|_{\mathfrak{V},2}^2 + &\lambda \, \| c \|_{\mathfrak{U}, 1} + \mu \, \| u\|_{\mathfrak{V},{\rm TV}} \Big\} \\ 
\mbox{s.t. } \| \mathcal{A}\{c\}  &- u \|_{\mathfrak{V},2}^2 = 0 \notag.
\end{align}  
Then, starting from an arbitrary $b^{0} \in \mathfrak{V}$, the Bregman algorithm \cite{Bregman:1967kx} finds optimal $c$ and $u$ through the following iterations
\begin{align}\label{Split-B}
\left( u^{t+1}, c^{t+1} \right) = \arg \min_{c, u}  \Big\{ \frac{1}{2} \| u - s \|_{\mathfrak{V},2}^2 + &\lambda \, \| c \|_{\mathfrak{U}, 1} + \mu \, \| u\|_{\mathfrak{V},{\rm TV}} + \frac{\gamma}{2} \| u -\mathcal{A}\{c\}  - b^{t} \|_{\mathfrak{V},2}^2 \Big\} \\
b^{t+1} = &b^{t} + \left( \mathcal{A}\{c^{t+1}\}  - u^{t+1} \right) \notag,
\end{align}
for some $\gamma > 0$\footnote{Note that the algorithm is guaranteed to converge for any $\gamma > 0$. In this work we use $\gamma = 0.5$.}. The functional in (\ref{Split-B}) is supposed to be minimized over two variables, i.e. $u$ and $c$. However, due to the way the $\ell_1$ and TV components of this functional have been split, the minimization can now be performed efficiently by iteratively minimizing with respect to $u$ and $c$ separately. The resulting iteration steps are
\begin{align}\label{steps}
\mbox{Step 1: } &c^{t+1} = \arg \min_{c}  \Big\{ \frac{\gamma}{2} \| \mathcal{A}\{c\}  - (u^{t} - b^{t}) \|_{\mathfrak{V},2}^2 +  \lambda \, \| c \|_{\mathfrak{U}, 1}  \Big\} \\
\mbox{Step 2: } &u^{t+1} = \arg \min_{u}  \Big\{ \frac{1}{2} \| u - s \|_{\mathfrak{V},2}^2 + \frac{\gamma}{2} \| u -(\mathcal{A}\{c^{t+1}\}  + b^{t}) \|_{\mathfrak{V},2}^2 + \mu \, \| u\|_{\mathfrak{V},{\rm TV}} \Big\} \notag.
\end{align}
Note that the functional at Step 2 contains two quadratic terms which can be combined together to result in
\begin{equation}\label{step2}
\mbox{Step 2: } u^{t+1} = \arg \min_{u}  \Big\{ \frac{1+\gamma}{2} \big\| u - \frac{s + \gamma \,(\mathcal{A}\{c^{t+1}\}  + b^{t}) }{1+ \gamma} \big\|_{\mathfrak{V},2}^2 + \mu \, \| u\|_{\mathfrak{V},{\rm TV}} \Big\}.
\end{equation}

To cause a substantial reduction in the value of the cost functional in (\ref{Split-B}), Step 1 and Step 2 should be applied recursively for a predefined number of times before the Bregman parameter $b^{t}$ is updated according to (\ref{Split-B}). It was argued in \cite{Goldstein:2008yo}, however, that the extra precision gained through such a repetitive application of Step 1 and Step 2 is likely to be ``wasted" when $b^{t}$ is updated. Consequently, it was suggested in \cite{Goldstein:2008yo} to perform these steps only once per iteration cycle. It is interesting to note that, in this case, the split Bregman algorithm transforms into the {\em alternating directions method of multipliers} (ADMM) \cite{Esser:2008jo}, whose convergence is guaranteed by the Eckstein-Bertsekas theorem \cite{Eckstein:1992vn} (see also Theorem 3.1 in \cite{Esser:2008jo}). 

The final algorithm is summarized below. Lines 3-4 of Algorithm 1 correspond to Step 1 in (\ref{steps}), while lines 5-6 correspond to Step 2. An even more important fact to notice is that the optimization problem in line 4 is separable in the spatial coordinate $\rr$. This optimization, therefore, can be performed at each voxel independently as discussed in Section~\ref{SP-only}. Moreover, the optimization problem in line 6 is separable in the diffusion coordinate $k$, and hence it amounts to applying TV-denoising to each of the $K$ components of $u$ independently as discussed in Section~\ref{TV-only}.

\begin{algorithm}
\caption{ADMM algorithm for sparse-TV reconstruction of HARDI signals} 
\begin{algorithmic}[1]
\STATE $b \Leftarrow 0$, $u \Leftarrow s$
\WHILE{``$c$ keeps changing"}
\STATE $d \Leftarrow u - b$
\STATE $c \Leftarrow \arg \min_{c}  \Big\{ \frac{1}{2} \| \mathcal{A}\{c\}  - d \|_{\mathfrak{V},2}^2 +  \frac{\lambda}{\gamma} \, \| c \|_{\mathfrak{U}, 1}  \Big\}$
\STATE $d \Leftarrow (1+\gamma)^{-1} \left(s + \gamma \, (\mathcal{A}\{c\}  + b)\right)$
\STATE $u \Leftarrow \arg \min_{u}  \Big\{ \frac{1}{2} \| u - d \|_{\mathfrak{V},2}^2 + \frac{\mu}{1+\gamma} \, \| u\|_{\mathfrak{V},{\rm TV}} \Big\}$
\STATE $b \Leftarrow b + \left( \mathcal{A}\{c\}  - u \right)$
\ENDWHILE
\end{algorithmic}
\end{algorithm}

%----------------------------------------------------- RESULTS ------------------------------------------------------
\begin{figure}[htbp]
\centering
\includegraphics[width=5in]{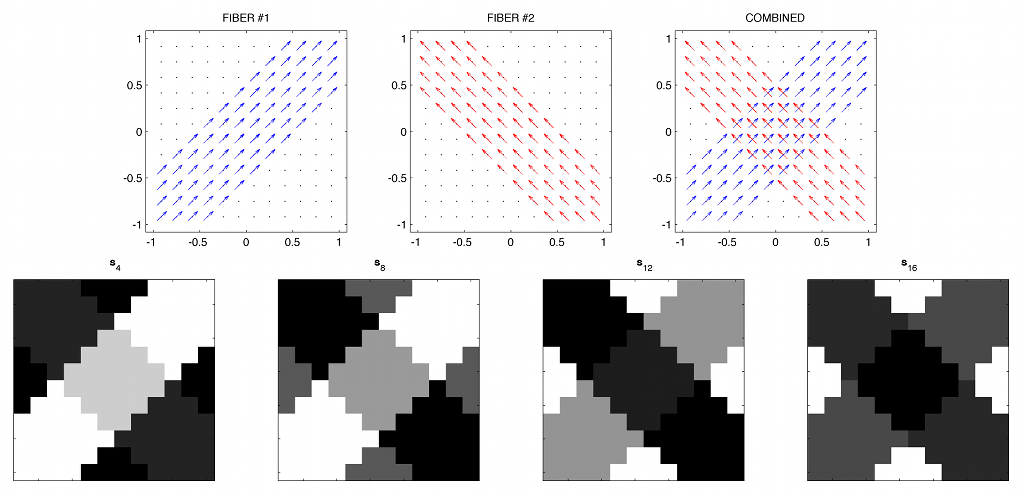} 
\caption{Phantom \#1: (Upper row of subplots) The orientations of the individual diffusion flows and their combination; (Lower row of subplots) Examples of the resulting (noise-free) diffusion-encoding images corresponding to four different diffusion-encoding directions.}
\label{F1}
\end{figure}

\begin{figure}[htbp]
\centering
\includegraphics[width=5in]{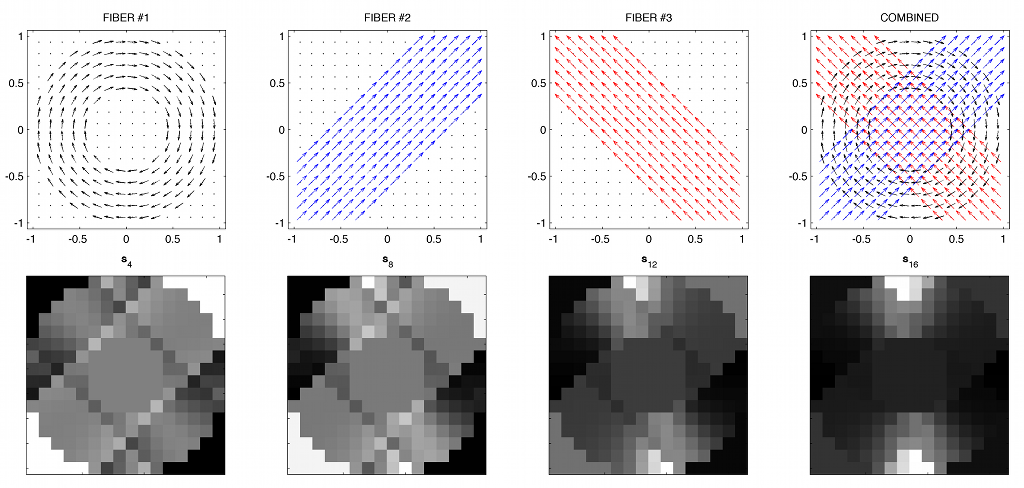} 
\caption{Phantom \#2: (Upper row of subplots) The orientations of the individual diffusion flows and their combination; (Lower row of subplots) Examples of the resulting (noise-free) diffusion-encoding images corresponding to four different diffusion-encoding directions.}
\label{F2}
\end{figure}

\begin{figure}[htbp]
\centering
\includegraphics[width=5in]{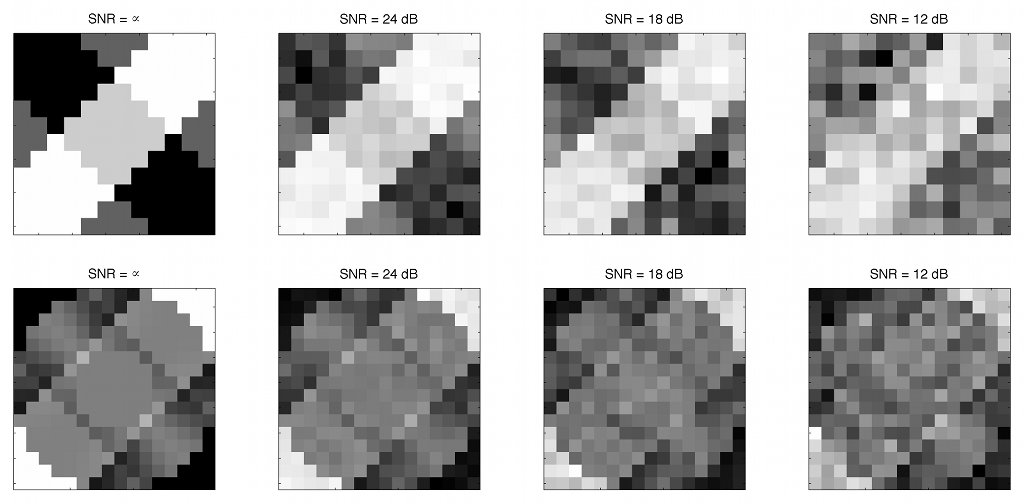} 
\caption{(Upper row of subplots) Diffusion-encoding images of Phantom \#1 corresponding to $\uu = [1, 1, 1] \slash \sqrt{3}$ and SNR = $\infty$, 24, 18 and 12 dB; (Lower row of subplots) Diffusion-encoding images of Phantom \#2 corresponding to the same $\uu$ and the same values of SNR.}
\label{F3}
\end{figure}

\section{Results}
\subsection{Technical details of the experimental study}
Both the choice of diffusion-encoding directions $\{\uu_k\}_{k=1}^K$ and of the orientations of spherical ridgelets require the use of a sampling scheme. In dMRI, one of the standard methods used to distribute a given number of spherical points over $\Es$ in a quasi-uniform manner is by means of the electrostatic repulsion algorithm (also known as the Thomson problem). However, since the diffusion signals are symmetric (implying $s(\uu \,|\, \rr) = s(- \uu \,|\, \rr)$), it is a unit hemisphere, not the entire $\Es$, which actually needs to be discretized. In view of the absence of a formulation of the Thomson problem for hemispherical domains, a common practice is to run the standard procedure for twice as many points as needed, followed by keeping only a half of the resulting configuration. However, as the retained points are not explicitly constrained to lie on a hemisphere, they may include nearly antipodal pairs which are likely to introduce undesirable dependencies between the diffusion measurements as well as between the basis functions. This limitation can be overcome through adapting a different sampling strategy. Particularly, in this paper, both the diffusion-encoding directions and the orientations of spherical ridgelets have been defined by using the method of generalized spiral points \cite{Rakhmanov:1994uq,Saff:1997ys}, in which the sampling points are arranged along a spiral in such a way that the distance between the points along the spiral is approximately equal to the distance between its coils. This method is easily adaptable for sampling of the ``northern" hemisphere (i.e. $\left\{ \uu \in \Es \,|\, \uu \cdot [0, \, 0, \, 1]^T \ge 0 \right\}$), providing a nearly uniform, unique and analytically computable coverage which is in no respect inferior to the one produced by solving the Thomson problem.  

To assess the performance of the proposed algorithm under controllable conditions, experiments with simulated data sets have been performed. In this case, the HARDI signals were generated according to model (\ref{multimodel}) with different values of $M(\rr)$, $D_i(\rr)$, and $s_0(\rr) = 1$, $\forall \rr$. The resulting signals were contaminated by variable levels of Rician noise, giving rise to a set of different SNRs. In this work, we adapt the standard definition of the SNR as
\begin{equation}\label{SNR}
{\rm SNR} = 20 \log_{10} \left( \frac{\| s - \tilde{s} \|_{\mathfrak{V},2}}{\| s \|_{\mathfrak{V},2}} \right), 
\end{equation}
where $s$ and $\tilde{s}$ denote an original signal and its noise-contaminated version, respectively, and the norms are computed as defined by (\ref{ell2}). It is worthwhile noting that the optimal values of regularization parameters $\lambda$ and $\mu$ in (\ref{L-cost}) are normally a function of the noise level. In the present paper, however, no attempts have been extended to optimize these values for different SNRs. Instead, it was found that $\lambda = 0.03$ and $\mu = 0.05$ provided acceptable estimation results in all the simulation scenarios, and hence these values have been used throughout the whole study. 

Following \cite{Michailovich:2010rp}, the scaling parameter $\rho$ in (\ref{kappaj}) was set to 0.5 and the resolution parameter $J$ in (\ref{ridges}) was set to be equal to 1, corresponding to a total of three resolution levels. The number of spherical ridgelet orientations were predefined with $m_0 = 4$, resulting in $M_{-1} = 16$, $M_0 = 49$ and $M_1 = 169$ ridgelets spanning the resolution levels $j=-1$, $j=0$ and $j=1$, respectively. Thus, the total number of spherical ridgelets used in the reconstruction was equal to 234.

To quantitatively compare the reconstruction results produced by the proposed and references methods for different numbers of sampling directions $K$ and various SNRs, three performance measures were used. The first of the three was the normalized mean-squared error (NMSE) defined as
\begin{equation}\label{NMSE}
{\rm NMSE} =  \frac{1}{N_x N_y N_z} \sum_{\rr \in \Omega_d} \frac{\| s(\rr) - \hat{s}(\rr) \|_2^2}{\| s(\rr) \|_2^2},
\end{equation}
with $s(\rr)$ being a reference HARDI signal corresponding to location $\rr$ and $\hat{s}(\rr)$ being its estimate. Depending on the nature of a specific experiment, the reference signal can be either a simulated signal discretized at 642 spherical points obtained by the 3rd order tessellation of the icosahedron {\em or} a signal reconstructed using a maximum possible number of diffusion-encoding orientations.

One of the most valuable outcomes of HARDI is in providing an access to computation of orientation distribution functions (ODFs) -- the functions whose modes are likely to coincide with the direction of local diffusion flows \cite{Tuch:2004cr}. Both the SH-based \cite{Descoteaux:2007jw} and ridgelet-based \cite{Michailovich:2010rp} methods of reconstruction of HARDI signals come with analytical expressions which relate the HARDI signals to their corresponding ODFs. The latter can in turn be used to recover the directions of local diffusion flows (or, equivalently, the orientations of their related fibre tracts) using, e.g., the steepest ascent procedure detailed in \cite{Michailovich:2010rp}. Suppose $\uu_0$ is the true direction of a diffusion flow and $\tilde{\uu}$ is its estimate. Then, the angular orientation error $\delta$ can be defined (in degrees) as
\begin{equation}\label{angerr} 
\delta = \frac{180}{\pi} \arccos (\uu_0 \cdot \tilde{\uu}).
\end{equation}
In this paper, as a performance measure, we use an {\em average} angular orientation error which is obtained by averaging the values of $\delta$ computed for all ``fibres" within a specified $\Omega_d$. 

The last performance measure used in this work is the probability $P_d$ of false fibre detection. To define $P_d$, let $M(\rr)$ be the true number of fibre tracts passing through voxel $\rr$ (as defined by model (\ref{multimodel})). Also, let $\hat{M}(\rr)$ be an estimated number of fibres, which is equal to the number of modes (maxima) of the ODF recovered at position $\rr$. Then, one can define
\begin{equation}\label{PD}
P_d = \left[ \frac{1}{N_x N_y N_z} \sum_{\rr \in \Omega_d} \frac{ | M(\rr) - \hat{M}(\rr) |}{M(\rr)}.\right] \cdot 100\%.
\end{equation}

In addition to the quantitative comparison, the reconstruction results will be evaluated through visual comparison as well. In this paper, we choose to visualize spherical functions by means of 3-D surface plots. Such a plot tends to project away from the origin of $\mathbb{R}^3$ in the directions along which a spherical function is maximized, while passing near the origin in the directions where the function approaches zero.

Finally, our choice of reference methods was motivated by the scope of the main statements made in this paper. First, since we argue that the frame of spherical ridgelets is optimally suited for CS-based reconstruction of HARDI signals, its performance has to be compared with that of alternative representation systems. In particular, the basis of spherical harmonics up to the order 8 inclusive has been used for a different definition of the sensing matrix $A$ in (\ref{matrix}). (Note that, in the case of a real and symmetric analysis, this SH-basis consists of 45 functions.) Additionally, the representation system proposed in \cite{Landman:2010pt,Landman:2010qq} has been exploited in the comparative study as well. This system is formed by applying a set of rotations to a Gaussian kernel of the form $d(\uu) = \exp \{ - b \, (\uu^T D_0 \, \uu)\}$, with $b$ defined as in (\ref{multimodel}) and $D_0$ equal to a (diagonal) diffusion tensor having a mean diffusivity of 766 mm$^2$/s and a fractional anisotropy of 0.8. Following \cite{Landman:2010pt}, the number of rotations (and hence the number of Gaussian basis functions) was set to be equal to 253. For the convenience of referencing, the CS-based reconstruction methods using the spherical ridgelets, the 8th-order spherical harmonics, and the rotated Gaussian kernels will be referred below to as the RDG, SH8 and GSS algorithms, respectively.

To assess the significance of the proposed spatial regularization, all the above algorithms have been applied with two different values of $\mu$ in (\ref{L-cost}), {\em viz.} $\mu = 0$ and $\mu = 0.05$. Note that, in the first of these cases, the spatial regularity is ignored, which leads to the sparse-only reconstruction discussed in Section~\ref{SP-only}. In the second case, on the other hand, the spatial regularity is taken into account and the reconstruction is performed by means of the split Bregman algorithm of Section~\ref{BREG}.

\begin{figure}[htbp]
\centering
\includegraphics[width=5in]{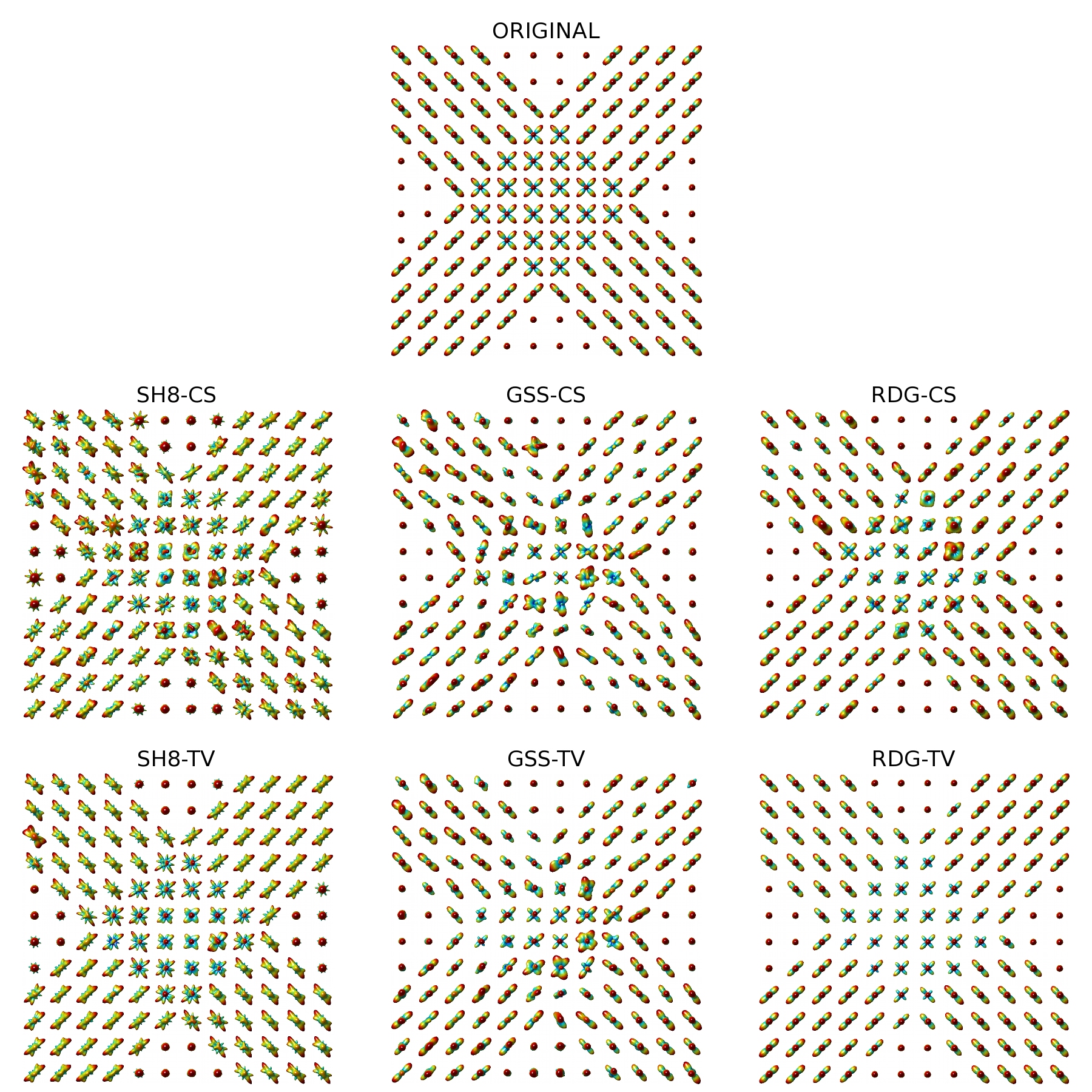} 
\caption{(Upper subplot) Original ODFs of Phantom \#1; (Middle row of subplots) The ODFs recovered by the SH8-CS, GSS-CS, and RDG-CS algorithms, respectively; (Bottom row of subplots) The ODFs recovered by the SH8-TV, GSS-TV, and RDG-TV algorithms, respectively.}
\label{F4}
\end{figure}

\begin{figure}[htbp]
\centering
\includegraphics[width=5in]{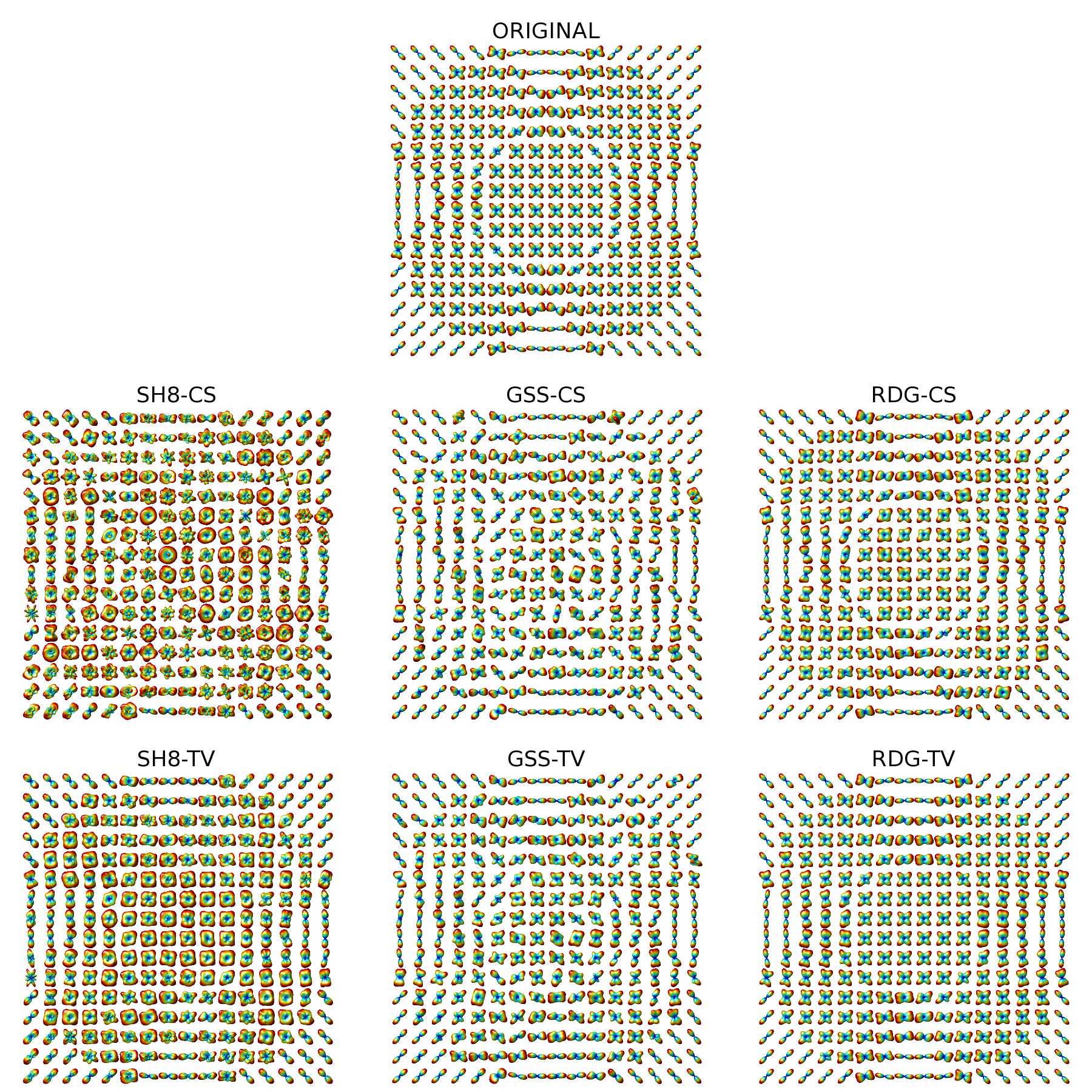} 
\caption{(Upper subplot) Original ODFs of Phantom \#2; (Middle row of subplots) The ODFs recovered by the SH8-CS, GSS-CS, and RDG-CS algorithms, respectively; (Bottom row of subplots) The ODFs recovered by the SH8-TV, GSS-TV, and RDG-TV algorithms, respectively.}
\label{F5}
\end{figure}

\subsection{In silico experiments}
To assess the performance of the proposed and reference methods under controllable conditions, two simulated data sets were used. The first set (referred to below as Phantom \#1) had a spatial dimension of $12\times 12$ pixels, and consisted of two ``fibres" crossing each other at the right angle as it is shown in the upper row of subplots of Fig.~\ref{F1}. In addition, each pixel in the set was assigned an extra diffusion flow in the direction perpendicular to the image plane. As a result, the number of diffusion components $M(\rr)$ in Phantom \#1 varied between 1 and 3. Subsequently, model (\ref{multimodel}) was used to generate corresponding diffusion-encode images $\{s_k\}_{k=1}^K$ for a range of different values of $K$. Two different values of $b$, namely $b = 1000$ s/mm$^2$ and $b = 3000$ s/mm$^2$ were used for data generation. The diffusion tensors $D_i(\rr)$ in (\ref{multimodel}) were obtained by applying rotations to a tensor of the form $D_0 = {\rm diag} \left( [ \alpha, \, \beta, \, \beta] \right)$, where $\alpha$ and $\beta$ were equal to $1700 \cdot 10^{-6}$ and $300 \cdot 10^{-6}$, respectively. Note that the mean diffusivity and fractional anisotropy of $D_0$ are equal to 766 mm$^2$/s and 0.8, respectively. Thus, the same diffusion tensors were used for data synthesis and for the construction of basis functions in the GSS algorithm, thereby allowing the latter to perform under the best possible conditions.

The lower row of subplots in Fig.~\ref{F1} depict four examples of the diffusion-encoding images obtained for Phantom~\#1 before their contamination by Rician noise. One can see that the images are piecewise constant functions, which appears to be in a good agreement with the bounded-variation model suggested by (\ref{L-cost}). However, real images may be more complicated than that. Accordingly, to test the robustness of the proposed regularization scheme, a different {\em in silico} phantom was designed. Phantom~\#2 had a spatial dimension of $16\times 16$ pixels and it was obtained through supplementing the configuration of Phantom~\#1 by an additional circular ``fibre" as shown in the upper row of subplots in Fig.~\ref{F2}. The lower row of subplots of the figure show a subset of the resulting diffusion-encoded images, which can be seen to no longer exhibit a piecewise constant behaviour characteristic for Phantom~\#1.

The simulated diffusion-encoded images were contaminated by three different levels of Rician noise, giving rise to SNR of 24, 18 and 12 dB. Some typical examples of the resulting images are demonstrated in Fig.~\ref{F3}, where the upper row of subplots depict a noise-free version of one of the diffusion-encoded images pertaining to Phantom~\#1 along with its noise-contaminated counterparts. The lower row of subplots in Fig.~\ref{F3} depict an analogous set of examples for Phantom~\#2. Observing the figure, one can see that the SNR values have been chosen so as to cover a range of possible noise scenarios, which could be characterized as moderate-to-severe contamination.

\begin{figure}[htbp]
\centering
\includegraphics[width=5in]{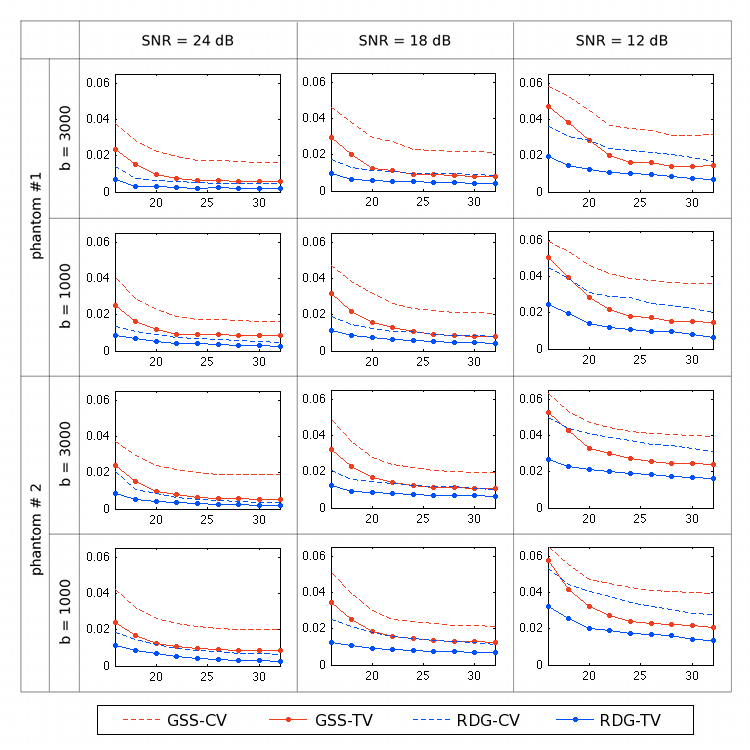} 
\caption{NMSE obtained using the compared methods for different phantoms, SNRs and $b$-values.}
\label{F6}
\end{figure}

\begin{figure}[htbp]
\centering
\includegraphics[width=5in]{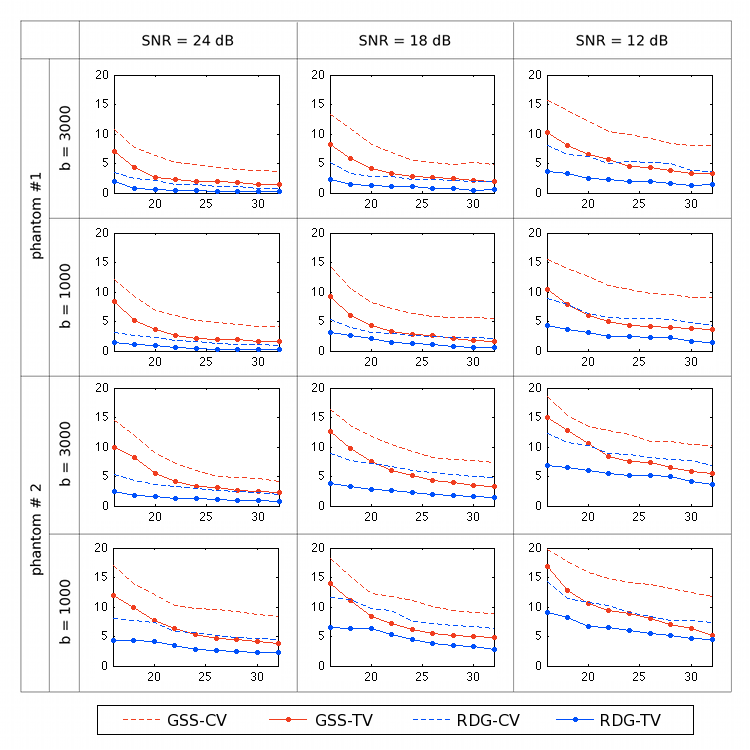} 
\caption{Average angular error $\delta$ obtained using the compared methods for different phantoms, SNRs and $b$-values.}
\label{F7}
\end{figure}

\begin{figure}[htbp]
\centering
\includegraphics[width=5in]{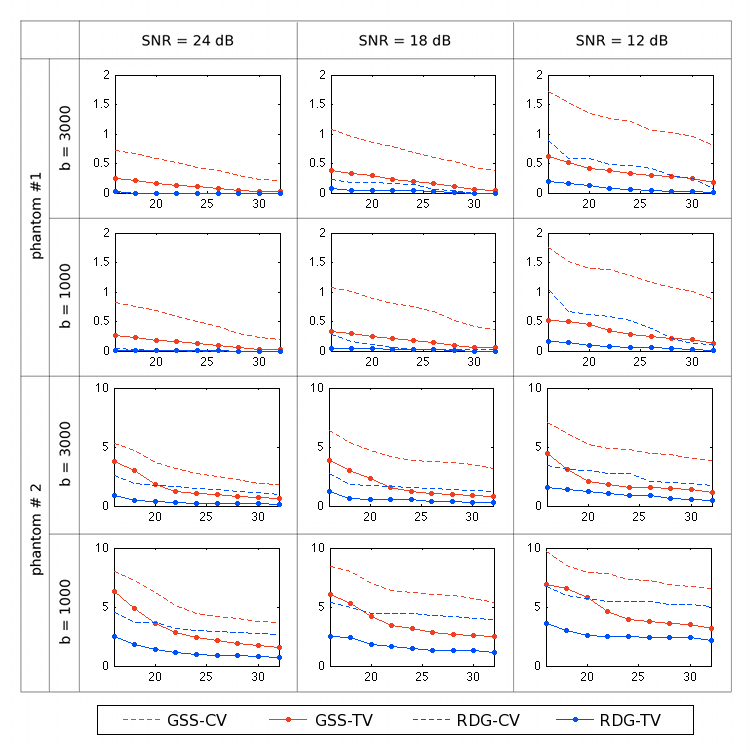} 
\caption{The rate of false fibre detection $P_d$ obtained using the compared methods for different phantoms, SNRs and $b$-values.}
\label{F8}
\end{figure}

As it was mentioned earlier, in our {\it in silico} study we compared the performance of three different representation bases, i.e. spherical harmonics (SH8), Gaussian kernels (GSS) and spherical ridgelets (RDG). All the resulting algorithms have been further subdivided into two different types, depending on whether or not the spatial regularization was engaged. Thus, in the absence of the spatial regularization (corresponding to $\mu = 0$), the reconstruction has been performed on a voxel-by-voxel basis, as detailed in Section~\ref{SP-only}. For the convenience of referencing, the corresponding algorithms will be referred to below as SH8-CS, GSS-CS, and RGD-CS. In the case of $\mu > 0$, the estimation has been carried out using the split Bregman method of Section~\ref{BREG}. The corresponding algorithms will be referred below as SH8-TV, GSS-TV, and RGD-TV.

The upper subplot of Fig.~\ref{F4} shows the original field of ODFs of Phantom~\#1 (corresponding to $b = 3000$ s/mm$^2$), which have been computed based on Tuch's approximation \cite{Tuch:2004cr} (i.e. by applying the Funk-Radon transform to the diffusion signals). At the same time, the middle row of subplots of Fig.~\ref{F4} show the ODFs recovered by (from left to right) SH8-CS, GSS-CS and RDG-CS with $K=16$ and SNR = 18 dB. One can see that the inability of the SH basis to sparsely represent HARDI signals results in a poor performance of SH8-CS. A better result is obtained with GSS-CS, which uses a basis of rotated Gaussian kernels, and therefore has a potential to represent the HARDI signals in a sparse manner. Unfortunately, the excessive correlation between the Gaussian basis functions adversely affects the ability of this method to withstand the effect of noise. Consequently, the reconstruction obtained using GSS-CV suffers from sizeable errors. The RDG-CS method, on the other hand, provides an estimation result of a much higher quality, albeit some inaccuracies are still noticeable in the central part of the phantom. The reconstruction accuracy improves dramatically when the spatial regularization is ``switched on", as it is demonstrated by the bottom row of subplots in Fig.~\ref{F4}. Specifically, while SH8-TV is still unable to provide a valuable reconstruction, the estimates obtained using GSS-TV and RDG-TV represent correctly the ``flow structure" of Phantom~\#1. Moreover, among the latter two methods, RDG-TV is clearly the best performer, resulting in a close-to-ideal recovery of the original ODFs. The superiority of RDG-TV over the alternative methods is further evident in the results presented by Fig.~\ref{F5}, which depicts the reconstructions obtained for Phantom~\#2 (with the same values of $b$, $K$ and SNR as above).

In general, the reconstruction results obtained using SH8-CS and SH8-TV have been observed to be of a lower quality in comparison to the other methods under consideration. For this reason, in what follows, only the GSS and RDG methods are compared. Thus, Fig.~\ref{F6} contrasts the performances of GSS-CS, GSS-TV, RDG-CS and RDG-TV in terms of the NMSE criterion. One can see that the best performance here is attained by the RDG-TV algorithm, which results in the smallest values of NMSE for both phantoms and for all the tested values of $b$, SNR and $K$. It is also interesting to note that the incorporation of spatial regularization allows GSS-TV to outperform RDG-CS, with the effect of the regularization becoming more pronounced at lower SNRs. On the whole, all the NMSE curves demonstrate an expected behaviour, with the error values increasing proportionally with a decrease in SNR, while going down with an increase in the number of diffusion-encoding gradients $K$. However, as opposed to the others, the NMSE curves obtained with RDG-TV are characterized by a relatively low rate of convergence, which indicates a reduced sensitivity of RDG-TV to the value of $K$.   

The above algorithms have been also compared in terms of the angular error (\ref{angerr}). The results of this comparison are summarized in Fig.~\ref{F7}, which again indicates that the most accurate reconstruction is obtained using the RDG-TV method. In general, the angular error tends to grow as SNR decreases and to converge to a minimum as $K$ increases. As opposed to the case of NMSE, however, there is an additional dependency of the angular error on the type of a phantom in use as well as on the $b$-value. In particular, the errors obtained for Phantom~\#2 are (on average) greater than those obtained for Phantom~\#1. This discrepancy is rooted in the fact that Phantom~\#2 has a more complex ``fibre structure" as compared to Phantom~\#1. In particular, while the ``fibers" of Phantom~\#1 are designed to cross each other at the right angle, the ``fibres" of Phantom~\#2 are allowed to decussate at much smaller angles, which makes them much harder to resolve. Moreover, this effect becomes more noticeable with a decrease in the $b$-value, which reduces the resolution of q-ball imaging. Finally, we notice that, on average, GSS-TV performs better than RDG-CS (though still worse than RDG-TV), which justifies the value of spatial regularization.

The comparison in terms of the rate of false fibre detection $P_d$ (\ref{PD}) was last in the line of our {\it in silico} performance tests; its results are shown in Fig.~\ref{F8}. One can see that, in the case of Phantom~\#1, RDG-TV yields a virtually zero false detection rate for both values of $b$, whereas the other methods result in considerably higher values of $P_d$ (mainly due to the detection of spurious local maxima in the estimated ODFs). The situation is different for Phantom~\#2, where all the compared methods yield sizeable errors (especially for $b=1000$ s/mm$^2$). However, in comparative terms, the most accurate reconstruction is still obtained by means of the proposed RDG-TV algorithm.

\subsection{In vivo results}
As the next validation step, experiments with real HARDI data were carried out. The proposed algorithm was tested on human brain scans acquired on a 3-Tesla GE system using an echo planar imaging (EPI) diffusion-weighted image sequence. A double echo option was used to suppress eddy-current related distortions. To improve the spatial resolution of EPI, an eight channel coil was used to perform parallel imaging by means of the ASSET technique with a speed-up factor of 2. The data were acquired using 51 gradient directions (quasi-uniformly distributed over the northern hemosphere) with $b=1000$ s/mm$^2$. In addition, eight baseline ($b0$) scans were acquired, averaged and used for normalization. The following scanning parameters were used: TR = 17000 ms, TE = 78 ms, FOV = 24 cm, $144 \times 144$ encoding steps, and 1.7 mm slice thickness. All scans had 85 axial slices parallel to the AC-PC line covering the whole brain.

\begin{figure}[htbp]
\centering
\includegraphics[width=5in]{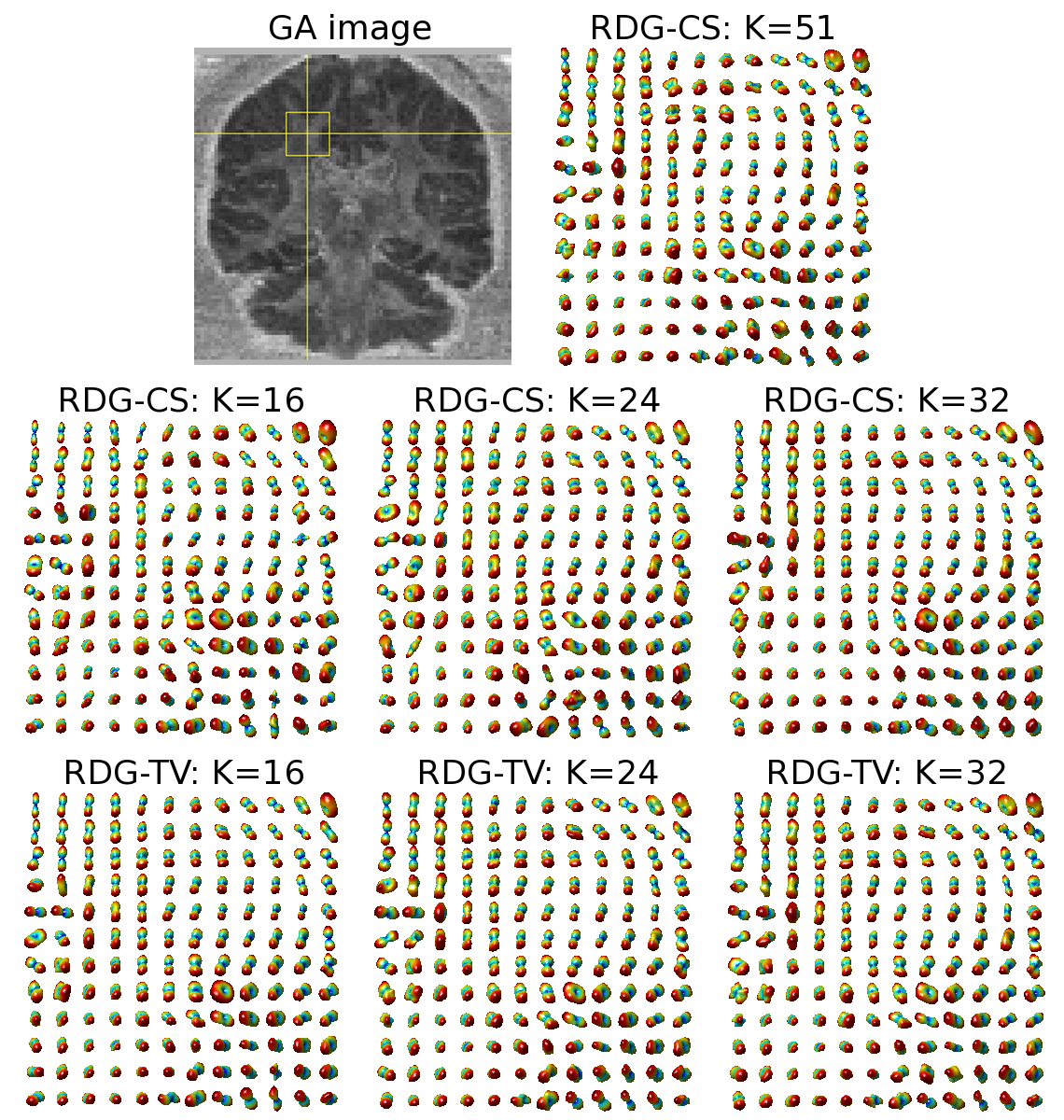} 
\caption{({\it Upper row of subplots}) A coronal GA image and the ODF field of the indicated region recovered by RDG-CS with $K=51$; ({\it Middle row of subplots}) Estimated ODF fields obtained using RDG-CS with $K=16$, $K=24$ and $K=32$; ({\it Bottom row of subplots}) Estimated ODF fields obtained using RDG-TV with $K=16$, $K=24$ and $K=32$.}
\label{F9}
\end{figure}

\begin{figure}[htbp]
\centering
\includegraphics[width=5in]{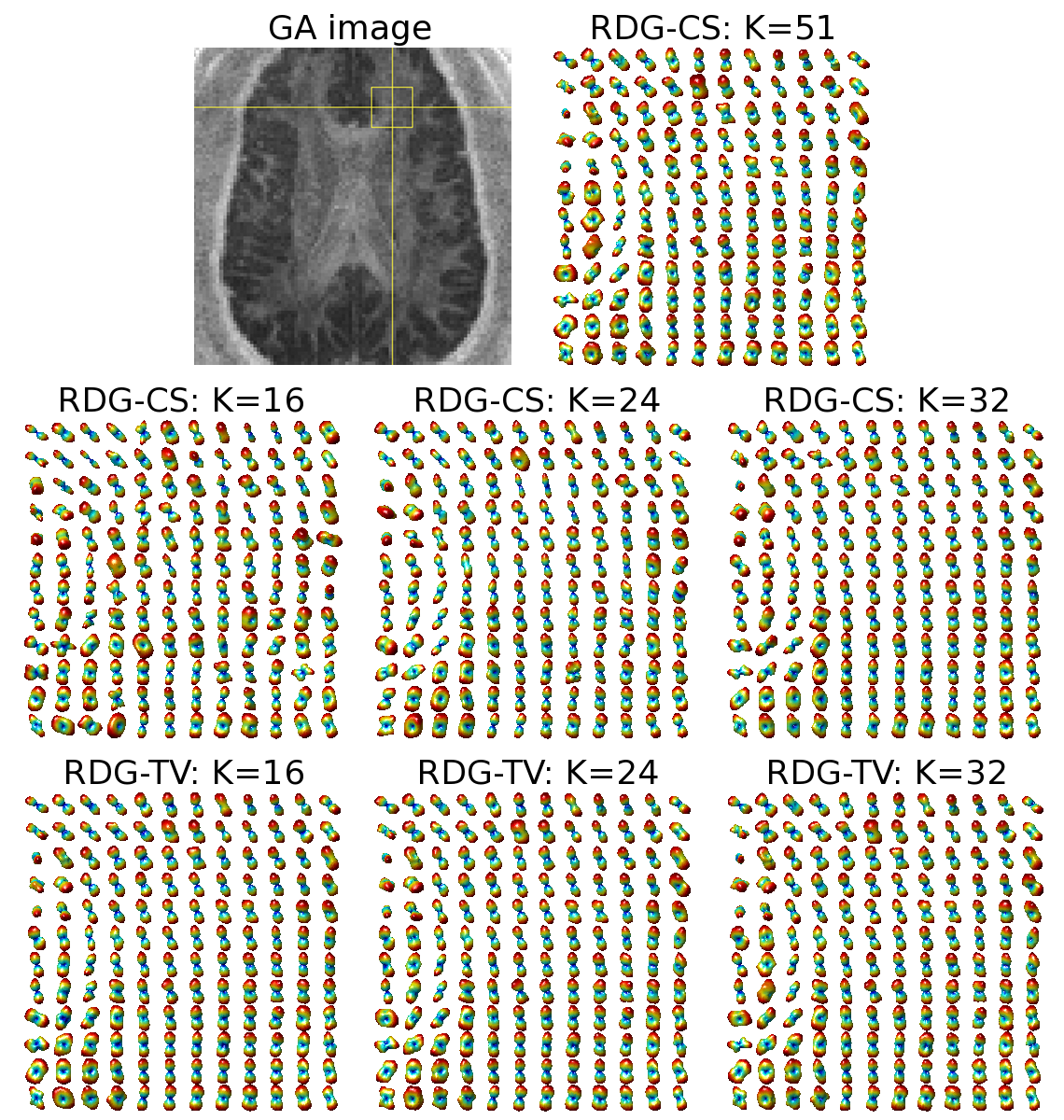} 
\caption{({\it Upper row of subplots}) An axial GA image and the ODF field of the indicated region recovered by RDG-CS with $K=51$; ({\it Middle row of subplots}) Estimated ODF fields obtained using RDG-CS with $K=16$, $K=24$ and $K=32$; ({\it Bottom row of subplots}) Estimated ODF fields obtained using RDG-TV with $K=16$, $K=24$ and $K=32$.}
\label{F10}
\end{figure}

The main question addressed through the {\it in vivo} experiments has been whether or not it is possible to supersede the spatial regularization by pre-filtering of HARDI signals. To this end, the RDG-CS algorithm was applied first to the HARDI data containing the full set of $K=51$ diffusion gradients. (Note that such {\em dense} reconstruction is analogous to the one reported in \cite{Michailovich:2010rp}, where the latter is shown to outperform the SH-based estimation \cite{Descoteaux:2007jw}.) The resulting ODFs have been used as a fiducial against which different reconstruction results were compared. 

As the next step, three different subsets of 16, 24 and 32 spherical points were composed out of the original set of 51 diffusion gradients. Within each of these subsets, their corresponding points were chosen so as to result in a quasi-uniform coverage of the northern hemisphere. Accordingly, the HARDI data were rearranged into three data sets of size $144\times144\times85\times16$, $144\times144\times85\times24$ and $144\times144\times85\times32$ to emulate compressed sensing data acquisition. The above sets were used to assess the performance of different reconstruction methods. Unfortunately, we have not succeeded to find conditions under which the SH8 and GSS algorithms would provide stable reconstruction results (either with or without pre-filtering). For this reason, only the RDG-CS and RDG-TV algorithms are compared below. 

\begin{table}[h]
\centering
\caption{NMSE computed between the dense and CS-based reconstructions obtained with RDG-CS and RDG-TV}
\begin{tabular}{| l | *{3}{c} | *{3}{c} |}
\hline
& \multicolumn{3}{c |}{Pertaining to Fig. 9} & \multicolumn{3}{c |}{Pertaining to Fig. 10} \\
\cline{2-7}
		& $K=16$ & $K=24$ & $K=32$ & $K=16$ & $K=24$ & $K=32$ \\
\hline
RDG-CS 	& 0.097	& 0.064	& 0.043	& 0.091	& 0.053	& 0.037 \\
RDG-TV	& 0.022	& 0.011	& 0.003	& 0.018	& 0.009	& 0.002 \\
\hline
\end{tabular}
\label{T1}
\end{table}

The upper row of subplots in Fig.~\ref{F9} show the generalized anisotropy (GA) \cite{Tuch:2004cr} image of a coronal cross-section of the brain along with the reference field of ODFs corresponding to the region indicated by the yellow rectangular. Anatomically, this region is expected to contain the fibre bundles of corona radiata as well as those of superior longitudinal and arcuate fasciculi. The middle row of subplots in the same figure depict the ODFs reconstructed by RDG-CS using $K=16, 24$ and $32$ diffusion gradients. One can see that the quality of reconstruction progressively improves as $K$ increases. It is important to note that, before applying the RDG-CS algorithm, the diffusion-encoded images had been pre-processed by a TV filter to reduce the effect of measurement noises on the estimation result. However, this pre-processing appears to be not nearly as effective as the spatial regularization of the RDG-TV algorithm, whose reconstruction results are shown in the bottom row of subplots in Fig.~\ref{F9}. The above conclusion is further supported by an additional example of Fig.~\ref{F10}, which shows the reconstructions pertaining to the indicated area within an axial cross-section of the brain. (The relevant fibre bundles here are those of cingulum and corpus callosum). As in the previous example, one can see that the most accurate reconstruction is attained by means of the proposed RDG-TV method. The superiority of RDG-TV is also confirmed by the quantitative figures of Table~\ref{T1}, which summarizes the NSME obtained by the compared algorithms for different values of $K$.

\section{Discussion and Conclusions}
When considered as a whole, the HARDI signals which pertain to a given volume of interest can be described as multi-valued (or, more generally, measure-valued) functions from a subset of $\mathbb{R}^3$ to the space of square-integrable spherical functions $\El(\Es)$. Such functions can be thought of as if they had two ``modes of variation" - one in the spatial and another in the diffusion domain. Although applying various inverse problems (a particular instance of which is addressed by the theory of CS) along the spatial and diffusion coordinates independently is not new to the community of medical imaging scientists, formulating a CS reconstruction problem in {\em both domains simultaneously} has not been proposed before. Accordingly, the present paper introduced the RDG-TV algorithm which exploits the above idea and can be used for reliable reconstruction of HARDI signals from as few as $K=16$ diffusion-encoded scans (as compared to 60-80 scans required by existing reconstruction tools). The algorithm exploits the fact that HARDI signals can be sparsely represented by spherical ridgelets in the diffusion domain, while their associated diffusion-encoded images have bounded variation in the spatial domain. Moreover, it has been shown experimentally that either using different representation bases or excluding the spatial regularization would result in considerably less accurate reconstruction results.   

At the practical level, the reconstruction is implemented based on the split Bregman approach (with 20 being the maximum number of Bregman iterations used in this study). The resulting algorithm alternates between two estimation stages (\ref{steps}): first, a sequence of basis pursuit de-noising problems are solved independently on a voxel-by-voxel basis, followed by applying a TV filter to a total of $K$ discrete images. Such computations are straightforward to accelerate using a multicore processing, which is another advantage of the proposed reconstruction method.

We believe that the algorithm presented in this paper can be improved in a number of ways. First, the square metric used to assess the model fidelity could be replaced by a different metric, which would be more specific to the nature of Rician noise. Second, the fact that diffusion signals are positive-valued could be explicitly incorporated into the reconstruction process in the form of additional constraints. Lastly, the bounded variation model could be substituted by an alternative model, which could (possibly) provide a better account for the spatial regularity of HARDI signals. Exploring the above options constitutes essential part of our ongoing research.

Finally, as the experimental study reported in this paper was comparative in its nature, it was not really important what method to use for approximation of ODFs. Specifically, the present results have been obtained using Tuch's approximation \cite{Tuch:2004cr}. However, more accurate computation of ODFs is possible based on the solid angle formulation as detailed in \cite{Aganj:2010fk,Tristan-Vega:2009uq}\footnote{We found, however, that the methods in \cite{Aganj:2010fk,Tristan-Vega:2009uq} are more sensitive to noise than the one in \cite{Tuch:2004cr}, and this is why we used the latter to avoid an inconsistency in interpreting the estimation results.}. It should also be noted that the technique proposed in \cite{Aganj:2010fk} can be applied to multi-shell HARDI data (i.e. the data acquired for a range of $b$-values). Until recently, collecting such data has been deemed impractical due to extremely long acquisition time required. We believe, however, that the proposed method for CS-based reconstruction of HARDI data has a potential to help multi-shell HARDI develop into a clinically relevant tool of diagnostic imaging.

\end{document}